\documentclass[prb, reprint, twocolumn, superscriptaddresss]{revtex4-1}
\def\be{\begin{equation}}
\def\ee{\end{equation}}
\def\bea{\begin{eqnarray}}
\def\eea{\end{eqnarray}}
\def\bi{\begin{itemize}}
\def\ei{\end{itemize}}
\usepackage{braket}

\usepackage{xcolor}
\usepackage{graphicx}
\usepackage{amsmath,amstext,amssymb,bm,color,times}
\usepackage[english]{babel}
\usepackage[T1]{fontenc}
\usepackage[utf8]{inputenc}
\usepackage[colorlinks=true,citecolor=blue,linkcolor=magenta]{hyperref}
\usepackage{lipsum}

\usepackage{comment}

\usepackage{tikz}
\usepackage{pgfplots}
\usepackage{adjustbox}
\usetikzlibrary{calc}

\begin{document}

\title{ Quantum Kibble-Zurek mechanism: \\
        Kink correlations after a quench in the quantum Ising chain } 

\author{Radosław J. Nowak}
\affiliation{Jagiellonian University, Institute of Theoretical Physics, {\L}ojasiewicza 11, PL-30348 Krak\'ow, Poland}
\author{Jacek Dziarmaga}
\affiliation{Jagiellonian University, Institute of Theoretical Physics, {\L}ojasiewicza 11, PL-30348 Krak\'ow, Poland}
\date{\today}

\begin{abstract}
The transverse field in the quantum Ising chain is linearly ramped from the para- to the ferromagnetic phase across the quantum critical point at a rate characterized by a quench time $\tau_Q$.  
We calculate a connected kink-kink correlator in the final state at zero transverse field.
The correlator is a sum of two terms: 
a negative (anti-bunching) Gaussian that depends on the Kibble-Zurek (KZ) correlation length only
and a positive term that depends on a second longer scale of length.
The second length is made longer by dephasing of the state excited near the critical point during the following ramp across the ferromagnetic phase. 
This interpretation is corroborated by considering a linear ramp that is halted in the ferromagnetic phase for a finite waiting time and then continued at the same rate as before the halt.
The extra time available for dephasing increases the second scale of length that asymptotically grows linearly with the waiting time. 
The dephasing also suppresses magnitude of the second term making it negligible for waiting times much longer than $\tau_Q$.
The same dephasing can be obtained with a smooth ramp that slows down in the ferromagnetic phase.
Assuming sufficient dephasing we obtain also higher order kink correlators and the ferromagnetic correlation function.
\end{abstract}
\maketitle

\section{ Introduction } 

Kibble-Zurek mechanism (KZM) originated from a scenario for defect creation in cosmological symmetry-breaking phase transitions~\cite{K-a, *K-b, *K-c}. As the Universe cools, causally disconnected regions must choose broken symmetry vacua independently resulting in topologically nontrivial configurations that survive as topological defects. In this Kibble scenario it is the speed of light that limits the size of the correlated domains.
In contrast a dynamical theory for the laboratory phase transitions \cite{Z-a, *Z-b, *Z-c, Z-d} employs equilibrium critical exponents of the transition and the quench time to predict the scaling of the resulting density of defects. KZM was successfully tested by numerical simulations \cite{KZnum-a,KZnum-b,KZnum-c,KZnum-d,KZnum-e,KZnum-f,KZnum-g,*KZnum-h,*KZnum-i,KZnum-j,KZnum-k,KZnum-l,KZnum-m,that} and laboratory experiments in condensed matter systems \cite{KZexp-a,KZexp-b,KZexp-c,KZexp-d,KZexp-e,KZexp-f,KZexp-g,KZexp-gg,KZexp-h,KZexp-i,KZexp-j,KZexp-k,KZexp-l,KZexp-m,KZexp-n,KZexp-o,KZexp-p,KZexp-q,KZexp-r,KZexp-s,KZexp-t,KZexp-u,KZexp-v,KZexp-w,KZexp-x}. More recently, KZM was adapted to quantum phase transitions \cite{QKZ1,QKZ2,QKZ3,d2005,d2010-a, d2010-b}. Theoretical developments \cite{QKZteor-a,QKZteor-b,QKZteor-c,QKZteor-d,QKZteor-e,QKZteor-f,QKZteor-g,QKZteor-h,QKZteor-i,QKZteor-j,QKZteor-k,QKZteor-l,QKZteor-m,QKZteor-n,QKZteor-o,KZLR1,KZLR2,KZLR3,QKZteor-q,QKZteor-r,QKZteor-s,QKZteor-t,sonic,QKZteor-u,QKZteor-v,QKZteor-w,QKZteor-x} as well and experimental tests \cite{QKZexp-a, QKZexp-b, QKZexp-c, QKZexp-d, QKZexp-e, QKZexp-f, QKZexp-g,deMarco2,Lukin18,adolfodwave} of the quantum KZM (QKZM) followed. The recent experiment~\cite{Lukin18}, where a quantum Ising chain in the transverse field is emulated with Rydberg atoms, is consistent with the theoretically predicted scalings~\cite{QKZ2,QKZ3,d2005}. 

In a cartoon version of QKZM, whose limitations --- but also essential correctness --- have been discussed in Ref.~\onlinecite{sonic}, the state of the system literally freezes-out in the neighborhood of the critical point due to the closing of the energy gap. In QKZM a system initially prepared in its ground state is smoothly ramped across a quantum critical point. A smooth ramp can be linearized near the critical point: 
\begin{equation}
\varepsilon(t)=\frac{t-t_c}{\tau_Q}.
\label{epsilont}
\end{equation}
Here $\varepsilon$ is a dimensionless parameter in the Hamiltonian, whose magnitude measures distance from the critical point, $\tau_Q$ is a quench time, and $t_c$ is the time when the critical point is crossed. Initially, far from the critical point, the evolution is adiabatic and the system follows its adiabatic ground state. The adiabaticity fails at $-\hat t$ when the reaction rate of the system, proportional the the gap $\Delta\propto|\varepsilon|^{z\nu}$, equals instantaneous transition rate $|\dot \varepsilon/\varepsilon| = 1/|t|$. Here $z$ and $\nu$ are, respectively, the dynamical and the correlation length exponent. From this equality we obtain 
\be 
\hat t\propto \tau_Q^{z\nu/(1+z\nu)}
\ee 
and the corresponding $\hat\varepsilon=\hat t/\tau_Q\propto\tau_Q^{-1/(1+z\nu)}$. In the cartoon ``freeze-out'' version of the impulse approximation the ground state at $-\hat\varepsilon$, with a corresponding correlation length,
\begin{equation}
\hat\xi \propto \tau_Q^{\nu/(1+z\nu)},  
\label{hatxi}
\end{equation}
is expected to survive until $+\hat t$, when the evolution can restart. In this way, $\hat\xi$ becomes imprinted on the initial state for the final adiabatic stage of the evolution after $+\hat t$. Oversimplified as it is, the adiabatic-impulse-adiabatic approximation predicts correct scaling of the characteristic lengthscale with $\tau_Q$, see Eq.~\eqref{hatxi}, and the timescale 
\be 
\hat t \propto \hat\xi^z.
\label{hatt}
\ee 
The post-quench density of excitations is determined by $\hat\xi$ within this scenario.  

In the integrable quantum ising chain the excitations are well defined as Bogoliubov quasiparticles \cite{d2005}. They get excited between $t_c-\hat t$ and $t_c+\hat t$. After $t_c+\hat t$, when the evolution of the system crosses over from the impulse to the adiabatic again, their power spectrum $p_k$ becomes frozen. Here $p_k$ is excitation probability for a pair of quasiparticles with opposite quasimomenta: $\pm k$. The excited state after $t_c+\hat t$ is a superposition over many eigenstates. Magnitudes of their amplitudes are determined by $p_k$, and thus remain frozen, but the amplitudes accumulate dynamical phases that depend on $k$. This may eventually lead to dephasing: the $k$-dependent phases become so scrambled that observables that are localized in space can be accurately calculated within an approximation of random phases \cite{QKZteor-e}.   

\begin{figure}
    \centering
    \includegraphics[width=1\linewidth]{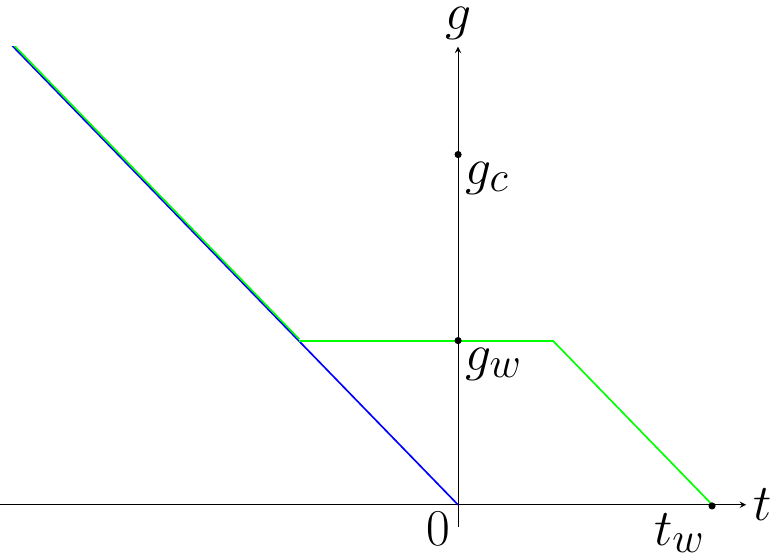}
    \caption{{\bf Transverse field ramp. }
Time dependence of the transverse magnetic field $g(t)$.
The blue line is the straight linear ramp that crosses the quantum critical point at $g_c=1$ before it stops at $g(0)=0$ where measurements are made. 
The green line is the ramp with a halt at $g_w<g_c$ for a waiting time $t_w$.
The halt allows for extra dephasing of quasiparticles that were excited in the Kibble-Zurek regime near the critical point: within $\pm\hat t$ from the time when the ramp crosses the critical point.}
    \label{fig:ramp}
\end{figure}

Motivated by new experimental opportunities opened by Rydberg atoms \cite{Lukin18,rydberg2d1,rydberg2d2,Semeghini2021}, in this paper we go beyond the set of observables considered in Ref. \onlinecite{QKZteor-e} and calculate correlations between ferromagnetic kinks at the end of the quench. This problem was first considered in Ref. \onlinecite{roychowdhury2020dynamics} in an elegant dual formulation of the quantum Ising chain. A similar problem was addressed in the 3D Kiatev model \cite{KZKitaev3D}. In addition to the probability distribution of the total number of kinks \cite{QKZteor-e,delCampo_distribution,adolfodwave}, these are experimentally most accessible predictions that go ``beyond KZM'', i.e., beyond the most basic average density of kinks. We obtain the kink-kink correlator in a closed analytic form without any random-phase approximation: the dephasing after a generic quench is far insufficient to justify the approximation. However, the dephasing has an impact on the correlator through a second length scale, $l$, that it makes longer than the basic KZ correlation length $\hat\xi$, though for a generic quench the correction is only logarithmic in $\tau_Q$. We demonstrate that $l$ can be made much longer by slowing the ramp after $t_c+\hat t$ to provide more time for dephasing, see Fig. \ref{fig:ramp}. When the extra time is long enough then the correlator becomes the same as in the random phase approximation. In this regime we can proceed further and derive a compact formula for higher order kink correlators and the ferromagnetic spin-spin correlation function. 


The paper is organized as follows. In section \ref{sec:QIM} we recall basic facts about the quantum Ising chain. Linear quench/ramp of the transverse magnetic field is defined in section \ref{sec:linearquench}. In section \ref{sec:LZ} the ramped quantum Ising model is mapped to the Landau-Zener problem. This is where we derive quadratic correlators for Jordan-Wigner fermions, identify the effects of dephasing and introduce the dephasing length $l$. The kink-kink correlator is worked out in section \ref{sec:kklinear}. It is shown to be a sum of two terms depending either on $\hat\xi$ or $l$. In order to substantiate the discussion of dephasing, in section \ref{sec:kkhalt} we make the linear ramp halt in the ferromagnetic phase for a variable waiting time $t_w$, see Fig. \ref{fig:ramp}. A more general analytic formula for the kink-kink correlator is obtained that depends on $t_w$ through a generalized dephasing length $l_w$. For long enough extra waiting time $l_w$ grows linearly with $t_w$ and the $l_w$-dependent term in the correlator decays like $l_w^{-1}$. Eventually the random phase approximation becomes accurate and the dephased correlator depends on the KZ length $\hat \xi$ only. In sections \ref{sec:higher} and \ref{sec:spinspin} we take advantage of the dephasing and work out higher order kink correlators and the ferromagnetic spin-spin correlator, respectively. Finally, we summarize in section \ref{sec:summary}. 


\section{ Quantum Ising chain }
\label{sec:QIM}

The transverse field quantum Ising chain is
\be
H~=~-\sum_{n=1}^N \left( g\sigma^x_n + \sigma^z_n\sigma^z_{n+1} \right)~.
\label{Hsigma}
\ee
Here we assume periodic boundary conditions:
$ 
\vec\sigma_{N+1}~=~\vec\sigma_1~.
$
In the thermodynamic limit, $N\to\infty$, the quantum critical point at $g=1$ separates the paramagnetic ($|g|>1$) from the ferromagnetic ($|g|$<1) phase. For definiteness we assume that $N$ is even. After the Jordan-Wigner transformation,
\bea
&&
\sigma^x_n~=~1-2 c^\dagger_n  c_n~, \\
&&
\sigma^z_n~=~
-\left( c_n+ c_n^\dagger\right)
 \prod_{m<n}(1-2 c^\dagger_m c_m)~,
 \label{JW}
\eea
introducing fermionic annihilation operators $c_n$, which satisfy anticommutation relations 
$\left\{c_m,c_n^\dagger\right\}=\delta_{mn}$ and 
$\left\{ c_m, c_n \right\}=\left\{c_m^\dagger,c_n^\dagger \right\}=0$,
the Hamiltonian (\ref{Hsigma}) becomes \cite{LSM1,LSM2}
\be
 H~=~P^+~H^+~P^+~+~P^-~H^-~P^-~.
\label{Hc}
\ee
Above
\be
P^{\pm}=
\frac12\left[1\pm\prod_{n=1}^N\sigma^x_n\right]=
\frac12\left[1~\pm~\prod_{n=1}^N\left(1-2c_n^\dagger c_n\right)\right]
\label{Ppm}
\ee
are projectors on the subspaces with even ($+$) and odd ($-$) numbers of $c$-quasiparticles and  
\bea
H^{\pm}~=~
\sum_{n=1}^N
\left( 
g c_n^\dagger  c_n - c_n^\dagger  c_{n+1} - c_{n+1}  c_n - \frac{g}{2} +{\rm h.c.} 
\right)~.
\label{Hpm}
\eea
are the corresponding reduced Hamiltonians. The $c_n$'s in $H^-$ satisfy
periodic boundary conditions, $c_{N+1}=c_1$, but the $c_n$'s in $H^+$
must be anti-periodic: $c_{N+1}=-c_1$. 

The parity of the number of $c$-quasiparticles is a good quantum number and the ground state has even parity for any non-zero value of $g$. Assuming that time evolution begins in the ground state, we can confine to the subspace of even parity. $H^+$ is diagonalized by a Fourier transform followed by a Bogoliubov transformation \cite{LSM1,LSM2}. The anti-periodic Fourier transform is  
\be
c_n~=~ 
\frac{e^{-i\pi/4}}{\sqrt{N}}
\sum_k c_k e^{ikn}~,
\label{Fourier}
\ee
where the pseudomomenta $k$ take half-integer values:
\be
k~=~
\pm \frac12 \frac{2\pi}{N},
\dots,
\pm \frac{N-1}{2} \frac{2\pi}{N}~.
\label{k}
\ee
It transforms the Hamiltonian into
\bea
H^+~&=&
\sum_k
\left\{
2(g-\cos k) c_k^\dagger c_k + \right. \nonumber\\
&&
\left. 
\left(
 c^\dagger_k c^\dagger_{-k}+
 c_{-k} c_k
\right)
\sin k
-g
\right\}~.
\label{Hck}
\eea
Diagonalization of $H^+$ is completed by the Bogoliubov transformation:
\be
c_k~=~
u_k  \gamma_k + v_{-k}^*  \gamma^\dagger_{-k}~,
\label{Bog}
\ee
where the Bogoliubov modes $(u_k,v_k)$ are eigenstates of stationary Bogoliubov-de Gennes equations:
\bea
\epsilon~ u_k &=& +2(g-\cos k) u_k+2\sin k v_k~,\nonumber\\
\epsilon~ v_k &=& -2(g-\cos k) v_k+2\sin k u_k~. \label{stBdG}
\eea
For each $k$ they have two eigenstates with eigenenergies $\epsilon=\pm\epsilon_k$, where  
\be
\epsilon_k~=~2\sqrt{(g-\cos k)^2+\sin^2 k}~.
\label{epsilonk}
\ee
The positive energy eigenstate,
$
(U_k,V_k)
$,
defines a fermionic quasiparticle operator $\gamma_k~=~U_k^* c_k + V_{-k} c_{-k}^\dagger$,
and the negative energy eigenstate, 
$
(U^-_k,V^-_k)=(-V_k,U_k)~ 
$,
defines $\gamma_k^-=(U_k^-)^*c_k+V_{-k}^-c_{-k}^\dagger=-\gamma_{-k}^\dagger$. 
After the Bogoliubov transformation, up to an additive constant the Hamiltonian is equivalent to
\be
H^+~=~
\sum_k \epsilon_k~
\gamma_k^\dagger \gamma_k~
\label{Hgamma}
\ee
but the projector $P^+$ in Eq.~(\ref{Hc}) implies that only states with even numbers of quasiparticles belong to the spectrum of $H$. 

With the quasiparticle dispersion \eqref{epsilonk} at the critical $g=1$ we obtain a linear dispersion, $\epsilon_k\approx 2|k|$, for small $|k|$ which implies the dynamical exponent $z=1$. On the other hand, for $k=0$ we have $\epsilon_0\propto |g-1|^1$ for a near-critical $g\approx1$ which implies $z\nu=1$ and the correlation length exponent $\nu=1$. Consequently, the KZ scales are
\be 
\hat t \propto \hat\xi \propto \sqrt{\tau_Q}.
\label{KZising}
\ee 

\section{ Linear quench }
\label{sec:linearquench}

We ramp the Hamiltonian across the quantum critical point by a linear quench
\be
g(t\leq0)~=~-\frac{t}{\tau_Q}~.
\label{linear}
\ee
with the characteristic quench time $\tau_Q$. The ramp crosses the critical point at $t_c=-\tau_Q$. For the universal features of the QKZM it is enough to assume that the ramp can be linearized near the critical point, with a slope $-1/\tau_Q$, but here we proceed with a solution of the analytically tractable fully linear ramp. The system is initially in its ground state at large initial value of $g\gg 1$, but as $g$ is ramped down to zero, the system gets excited from its instantaneous ground state and, in general, its final state at $t=0$ has finite number/density of kinks. Comparing the Ising Hamiltonian Eq.~(\ref{Hsigma}) at $g=0$ with the Bogoliubov Hamiltonian (\ref{Hgamma}) at $g=0$ we obtain a simple expression for the operator of the total number of kinks 
\bea
{\cal N} ~=~ \sum_{n=1}^{N} K_n ~=~ \sum_k \gamma_k^\dagger \gamma_k~.   
\label{calN}
\eea 
Here
\be 
K_n = \frac12 \left( 1 - \sigma_{n}^z\sigma_{n+1}^z \right)
\label{K}
\ee
with eigenvalues $0,1$ is the kink number operator on the bond between sites $n$ and $n+1$. The total number of kinks is equal to the number of quasiparticles excited at $g=0$. 

\section{ Landau-Zener problem }
\label{sec:LZ}

The initial ground state is the Bogoliubov vacuum $|0\rangle$ annihilated by all quasiparticle operators $\gamma_k$. For the initial $g\gg1$ they are defined by the stationary Bogoliubov modes $(U_k,V_k)\approx(1,0)$. We assume the Heisenberg picture, where Fock states with definite quasiparticle occupations numbers do not change and, therefore, the Bogoliubov quasiparticle operators $\gamma_k$ expressed through these states do not change either but the Jordan-Wigner fermions evolve with the usual Heisenberg equation: $i\frac{d}{dt}c_k=[c_k,H^+]$. With a time-dependent Bogoliubov transformation,
\be
c_k ~=~ u_k(t)  \gamma_k + v_{-k}^*(t)  \gamma^\dagger_{-k},
\label{tildeBog}
\ee
the Heisenberg equation becomes equivalent to time-dependent Bogoliubov-de Gennes equations (\ref{stBdG}):
\bea
i\frac{d}{dt} u_k &=&
+2\left[g(t)-\cos k\right] u_k +
 2 \sin k~ v_k~,\nonumber\\
i\frac{d}{dt} v_k &=&
-2\left[g(t)-\cos k\right] v_k +
 2 \sin k~ u_k~.
\label{dynBdG} 
\eea
with the initial/asymptotic condition $[u_k(-\infty),v_k(-\infty)]=(1,0)$. 
They can be solved exactly by mapping to the Landau-Zener (LZ) problem\cite{d2005,QKZteor-e}. Indeed, a transformation to a new time variable:
\be
\tau~=~4\tau_Q\sin k\left(\frac{t}{\tau_Q}+\cos k\right)
\label{tau}
\ee
brings Eqs.~(\ref{dynBdG}) to the standard LZ form:
\bea
i\frac{d}{d\tau} u_k &=&
-\frac12(\tau\Delta_k) u_k + \frac12 v_k ~,\nonumber\\
i\frac{d}{d\tau} v_k &=&
+\frac12(\tau\Delta_k) v_k + \frac12 u_k ~,\label{LZ}
\label{BdGLZ}
\eea
with an effective transition rate $\Delta_k=(4\tau_Q\sin^2 k)^{-1}$. Here the new time $\tau$ runs from $-\infty$ to $\tau_{\rm final}=2\tau_Q\sin(2k)$ that corresponds to $t=0$. 

\subsection{ Spectrum of excitations and density of kinks }
\label{sec:pk}

For slow enough transitions only modes with small $k$, which have small gaps at their anti-crossing points, can get excited. For these modes $\tau_{\rm final}$ is much longer than the time when the anti-crossing is completed, $\propto \Delta_k^{-1/2}$, and we can use the LZ excitation probability:
\be
p_k~\approx~
e^{-\frac{\pi}{2\Delta_k}}~\approx~
e^{-2\pi\tau_Qk^2}~.
\label{LZpk}
\ee 
The approximations are accurate when $\tau_Q\gg1$. We can calculate the number of kinks in Eq.~(\ref{calN}) as
$ 
{\cal N}~=~\sum_k~p_k~.
$
In the thermodynamic limit, $N\to\infty$, the sum can be replaced by an integral and the density of kinks becomes
\cite{d2005}:
\be
n=\lim_{N\to\infty}\frac{\cal N}{N}=
\frac{1}{2\pi}\int_{-\pi}^{\pi}dk~p_k=
\frac{1}{2\pi\sqrt{2\tau_Q}}.
\label{scaling}
\ee    
The density scales like $\hat\xi^{-1}\propto\tau_Q^{-1/2}$ in agreement with QKZM. With kink correlations in mind it is natural to make definition of $\hat\xi$ precise as 
\be 
\hat\xi \equiv 2\pi\sqrt{2\tau_Q} = \frac{1}{n}.
\label{hatxiprecise}
\ee 
An inverse of this KZ length is equal to average final density of kinks/excitations.

\subsection{ Exact solution }
\label{sec:Weber}

The kink correlations will require more than just the excitation spectrum \eqref{LZpk}. A general solution of equations (\ref{BdGLZ}) is\cite{Damski_PRA_2006,QKZteor-e}: 
\bea
v_k(\tau)&=&
-\left[ a D_{-s-1}(-iz) + b D_{-s-1}(iz) \right] ~, \nonumber\\
u_k(\tau)&=&
\left(-\Delta_k\tau+2i\frac{\partial}{\partial\tau}\right)
v_k(\tau) ~,
\label{general} 
\eea
with free complex parameters $a,b$. Here $D_m(x)$ is a Weber function, 
$s=\frac{1}{4i\Delta_k}$, and $z=\sqrt{\Delta_k}\tau e^{i\pi/4}$.
The free parameters can be fixed by the asymptotic conditions: 
$u_k(-\infty)=1$ and $v_k(-\infty)=0$. Using the asymptotes of the Weber
functions when $\tau\to-\infty$, we obtain $a=0$ and 
\be
|b|^2=\frac{e^{-\pi/8\Delta_k}}{4\Delta_k}~.
\ee 
The exact solution of the linear quench problem is then
\bea
v_k(\tau)&=&
- b D_{-s-1}(iz) ~, \nonumber\\
u_k(\tau)&=&
\left(-\Delta_k\tau+2i\frac{\partial}{\partial\tau}\right)
v_k(\tau) ~,
\label{solution} 
\eea
At the end of the quench for
$t=0$ and when $\tau=\tau_k=2\tau_Q \sin(2k)$, the argument of the Weber function
$iz=\sqrt{\Delta_k}\tau e^{i\pi/4}=2\sqrt{\tau_Q}e^{i\pi/4}\cos(k){\rm sign}(k)$.

\subsection{ Fermionic correlators }
\label{sec:alphabeta}

For the considered $\tau_Q\gg1$ the magnitude of $iz$ is large for most $k$, except the neighborhoods of $k=\pm\frac{\pi}{2}$, and we can again use the asymptotes of the Weber functions to obtain \cite{QKZteor-e}
\bea
|u_k|^2&=&
\frac12(1-\cos k)+p_k~,\nonumber\\
u_k v_k^* &=&
\frac12\sin k+
{\rm sign}(k)~
\sqrt{p_k(1-p_k)}~
e^{i\varphi_k}~,\nonumber\\
\varphi_k &=&
\frac{\pi}{4}+
2\tau_Q-
(2-\ln 4)k^2\tau_Q+ \nonumber\\
&&
k^2 \tau_Q\ln\tau_Q-
\arg\left[\Gamma\left(1+ik^2\tau_Q\right)\right]~.
\label{att0smallk}
\eea
Here $\Gamma(x)$ is the gamma function and $\varphi_k$ is a dynamical phase acquired by a pair of excited quasiparticles with quasimomenta $(k,-k)$. These formulas depend on $k$ and $\tau_Q$ through two combinations: $\tau_Q k^2$, which implies the usual KZ correlation length, $\hat\xi\propto\tau_Q^{1/2}$, and $k^2 \tau_Q\ln\tau_Q$ which implies a second scale of length $\propto\sqrt{\tau_Q\ln\tau_Q}$. The final quantum state at $g=0$ cannot be fully characterized by a single scale of length. Physically, this reflects a combination of two processes: KZM that sets up the post-transition spectrum of excitations, $p_k$, and subsequent dephasing of the excited quasiparticle modes that manifests through the dynamical phase $\varphi_k$.

In order to make the phase more intelligible we can approximate ${\rm arg} [\Gamma \left(1+i\tau_Qk^2\right)]\approx -\gamma_E \tau_Qk^2$ for small enough $\tau_Qk^2$, where $\gamma_E$ is the Euler gamma constant. Given that excited quasiparticles have at most $\tau_Qk^2\approx 1/2\pi$, see \eqref{LZpk}, this is an accurate approximation that renders $\varphi_k$ quadratic in $k$:
\bea
\varphi_k-\varphi_0 &=& \left( \ln\tau_Q+\ln 4-2+\gamma_E \right) k^2\tau_Q \nonumber\\
                    &=& \left( \ln\tau_Q-0.036 \right) k^2\tau_Q \nonumber \\
            & \approx & k^2\tau_Q\ln\tau_Q.
            \label{quadraticvarphi}
\eea  
It also makes manifest that the dynamical phase is characterized solely by the second scale $\propto\sqrt{\tau_Q\ln\tau_Q}$.


\begin{figure}
    \centering
    \includegraphics[width=1\linewidth]{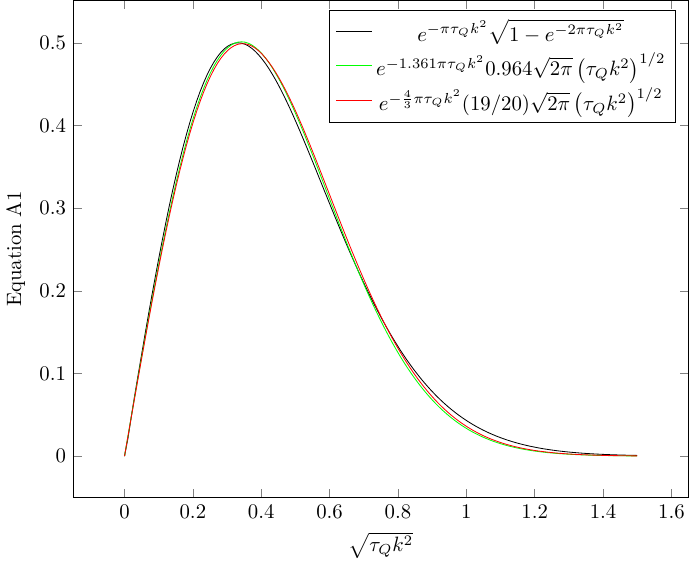}
    \caption{{\bf Approximation in Eq. \eqref{A1}. }
Comparison between the exact formula in Eq. \eqref{A1} (black) and the approximate one with either the raw parameters from minimization of the quadratic error: $A=0.964,a=1.361$ (green) or the adjusted ones: $A=19/20,a=4/3$ (red). 
}
    \label{fig:AB}
\end{figure}

The Gaussian state can be fully characterized by its two quadratic fermionic correlators:
\bea
\alpha_R &\equiv& \langle c_{n+R} c_n^\dagger \rangle = 
\frac{1}{2\pi}\int_{-\pi}^\pi dk~|u_k|^2~e^{ikR}~
   =   \nonumber\\
&& \frac12\delta_{0,|R|}-\frac14\delta_{1,|R|}+
   \hat\xi^{-1} e^{-\pi(R/\hat\xi)^2}~,
\label{alpha}
\eea
and 
\bea
\beta_R~\equiv~\langle c_{n+R} c_n \rangle = 
\frac{1}{\pi}\int_{0}^\pi dk~u_kv_k^*~\sin kR.
\label{beta}
\eea
With \eqref{att0smallk} we obtain 
\be 
\beta_R=\frac14 {\rm sign}(R)\delta_{|R|,1}+\delta\beta_R,
\label{betaR}
\ee 
where the first term is a ground state contribution while the second one comes solely from the excitations:
\be 
\delta\beta_R=
\frac{1}{\pi}\int_0^\pi dk~
e^{-\pi\tau_Q k^2}
\sqrt{1-e^{-2\pi\tau_Q k^2}}~
e^{i\varphi_k}
\sin kR.
\label{deltabeta}
\ee
In order to make the integral analytically tractable we make an approximation: 
\bea
e^{-\pi\tau_Q k^2} 
\sqrt{1-e^{-2\pi\tau_Q k^2}} \approx
e^{-a\pi\tau_Q k^2}
A \sqrt{2 \pi} 
\left(\tau_Qk^2\right)^{1/2}  .
\label{A1}
\eea
With $A=1$ and $a=1$ this would be just the leading term in a series expansion in powers of $\left(\tau_Qk^2\right)^{1/2}$. However, for greater accuracy we treat $A$ and $a$ as variational parameters. Minimization of the quadratic error of the approximation yields $A=0.964$ and $a=1.361$. Within its broad minimum we slightly adjust these numbers to
\be 
A=19/20, ~~~ a=4/3.
\ee
Comparisons between these two approximations and the exact formula are made in Fig. \ref{fig:AB}. 

Putting together all the approximations Eq. \eqref{deltabeta} becomes:
\be 
\delta\beta_R=
\frac{e^{i\varphi_0}\sqrt{2}A}{\sqrt{\pi\tau_Q}} 
\int_0^{\pi\sqrt{\tau_Q}} qdq~
e^{-a\pi q^2+i q^2 \ln\tau_Q} 
\sin\frac{q R}{\sqrt{\tau_Q}}.
\label{deltabetaapp0}
\ee 
Here $q=\sqrt{\tau_Q}k$.
After the upper limit of the integral is safely extended to infinity we obtain
\bea 
\delta\beta_R &=&
\frac{\sqrt{8\pi}A}{a^{3/2}}
\frac{R}{\sqrt{\hat\xi l^3}}
e^{ -\frac{2\pi}{a} (R/l)^2 }
e^{i\phi_R} \nonumber\\
&=&
\frac{57\sqrt{6\pi}}{80}
\frac{R}{\sqrt{\hat\xi l^3}}
e^{ -\frac{3\pi}{2} (R/l)^2 }
e^{i\phi_R},
\label{deltabetaapp}
\eea 
where 
$
\phi_R=
\frac14\pi+2\tau_Q-\frac32 {\rm arg}\left(1-\frac{3i\ln\tau_Q}{4\pi}\right)
-\frac98(R/l)^2\ln\tau_Q
$ 
is a phase and the correlation range is
\be 
l  = 
\hat\xi ~ \sqrt{ 1 + \left(\frac{3\ln\tau_Q}{4\pi}\right)^2 }.
\label{l}
\ee 
For very slow quenches, when $\ln\tau_Q\gg4\pi/3$, the range of this correlator becomes much longer than $\hat\xi$.

\begin{figure}[t!]
    \centering
    \includegraphics[width=1\columnwidth]{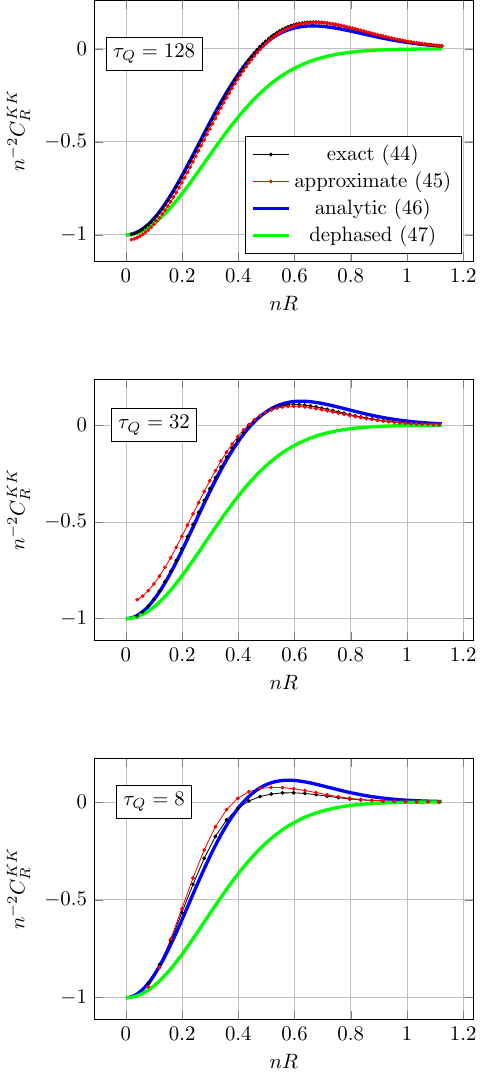}
  \caption{{\bf Kink-kink correlator. } 
Comparison between (scaled) exact correlator (\ref{CR}) and (scaled) approximate one (\ref{CRpm}), both plotted in function of scaled distance $n R$, demonstrating that the approximate one is accurate for long enough $\tau_Q$. We also plot the analytic formula in Eq. \eqref{scaledCR} to demonstrate that it becomes accurate for long enough $\tau_Q$.
The dephased correlator \eqref{CRpmw}, after the halted ramp with long enough waiting time, is also shown for comparison.
}
\label{fig:scaledCR}
\end{figure}
\section{Kink-kink correlator after a linear ramp}
\label{sec:kklinear}

The connected kink-kink correlator is
\bea 
C_{R}^{KK} &=& \langle K_{n} K_{n+R} \rangle_c \nonumber\\
           &=&
               \langle K_{n} K_{n+R} \rangle 
             - \langle K_{n} \rangle \langle K_{n+R} \rangle~,
\eea 
where $K_n$ is the kink number operator on the bond between sites $n$ and $n+1$, see \eqref{K}. In terms of the fermionic correlators it becomes 
\bea
 C_{R}^{KK} 
 &=&
 {\rm Re}\beta_{R+1} {\rm Re}\beta_{R-1} + \left( {\rm Im}\beta_R \right)^2 - \alpha_{R+1}\alpha_{R-1} + \nonumber\\
 &&
 \alpha_{R-1}{\rm Re}\beta_{R+1}  -\alpha_{R+1}{\rm Re}\beta_{R-1} ~.
 \label{CR}
\eea
When $R\pm 1$ is approximated by $R$, which should be accurate for the assumed $\hat\xi\gg1$, the correlator reduces to:
\be 
C_{R}^{KK} = \lvert \beta_{R} \rvert^2 - \alpha_{R}^2~.
\label{CRpm}
\ee 
Interestingly, this is $1/4$ of the connected transverse correlator:
$
C_R^{xx}=
\langle \sigma^x_n \sigma^x_{n+R} \rangle -
\langle \sigma^x_n     \rangle
\langle \sigma^x_{n+R} \rangle.
$
In order to properly assess the strength of kink-kink correlations, the correlator \eqref{CRpm} should be normalized by the square of the average density of kinks in (\ref{scaling}). After this normalization the exact formula in \eqref{CR} and the approximate one in \eqref{CRpm} are compared in Fig. \ref{fig:scaledCR}. As expected, they become the same for large enough $\tau_Q$. 

With equations \eqref{alpha}, \eqref{betaR}, \eqref{deltabetaapp}, and \eqref{CRpm} we obtain a compact analytic formula:
\be 
n^{-2} C^{KK}_R = 
\alpha~
\frac{\hat\xi}{l}
\left(\frac{R}{l}\right)^2 
e^{ -3\pi (R/l)^2 } 
- e^{ - 2\pi (R/\hat\xi)^2 }.
\label{scaledCR}
\ee 
Here $\alpha=\frac{ 9747 \pi }{3200}=9.57$ is a numerical prefactor. In figure \ref{fig:scaledCR} we compare the normalized correlator in \eqref{CRpm} with the analytic formula in \eqref{scaledCR} finding good agreement that is improving with increasing $\tau_Q$. The normalized correlator is of the order of $1$ implying strong correlation effects. Especially its second negative term implies strong anti-bunching. The kinks can hardly approach one another closer than a half of $\hat\xi$, i.e., half of the typical distance between them.


\begin{figure}
    \centering
    \includegraphics[width=1\columnwidth]{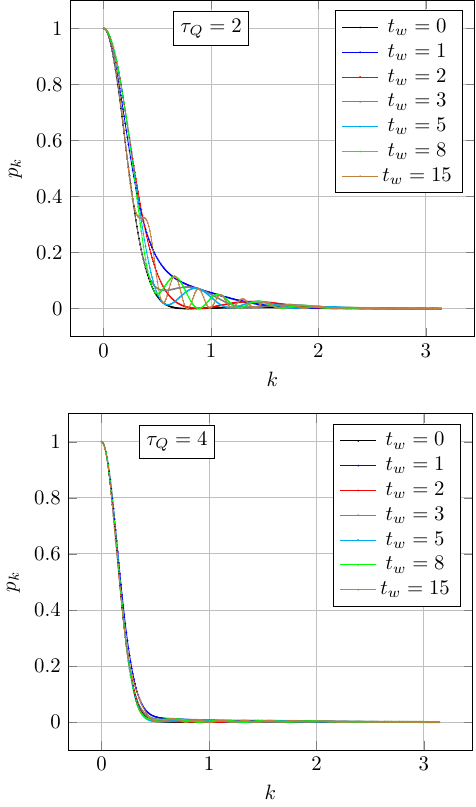}
    \caption{{\bf Power spectrum after the halt. }
Power spectrum $p_k=\langle \gamma_k^\dag \gamma_k \rangle$ at the final transverse field $g=0$ after a linear ramp with a halt at $g_w=1/2$ for a waiting time $t_w$, see Fig. \ref{fig:ramp}. As predicted, with increasing $\tau_Q$ the influence of discontinuous time derivative at the beginning and the end of the halt on the power spectrum quickly becomes negligible, compare $\tau_Q=2$ in the top panel with $\tau_Q=4$ in the bottom one. 
}
\label{fig:halt}
\end{figure}
\section{ Kink-kink correlator after a linear ramp with a halt }
\label{sec:kkhalt}

We have seen that there is an interplay between the KZ mechanism and the dephasing after $+\hat t$. Its manifestation are the two scales of length, $\hat\xi$ and $l$, that show up in the kink-kink correlator (\ref{scaledCR}). In order to make the distinction between the two effects even sharper, here we consider the same linear ramp as before but with an additional halt at $g_w\in(0,1)$ for a waiting time $t_w$, see Fig. \ref{fig:ramp}. We expect that for long enough waiting time the nontrivial quasiparticle dispersion (\ref{epsilonk}) will completely dephase excited quasiparticles with different quasimomenta. The dynamical phase $\varphi_k$ in \eqref{deltabeta} will depend on $k$ strongly enough for the magnitude of $\delta\beta_R$ to be suppressed and the kink-kink correlator to become
\be 
n^{-2} C^{KK}_R =
-e^{ -2\pi (R/\hat\xi)^2 }~.
\label{CRpmw}
\ee 
This purely negative dephased correlator demonstrates strong anti-bunching effect.

Formula \eqref{CRpmw} coincides with the one advocated in Ref. \onlinecite{roychowdhury2020dynamics} but, contrary to Ref. \onlinecite{roychowdhury2020dynamics}, waiting at the final $g_w=0$ cannot dephase the kink correlator because the dispersion (\ref{epsilonk}) at $g=0$ is flat: $\epsilon_k = 2$. Even if it were not, the kink number operator (\ref{K}) commutes with the Hamiltonian at $g=0$ and the correlator must remain constant there.   

The halt at $g_w>0$ is convenient analytically but it has a disadvantage that the discontinuous time derivative of the function $g(t)$ at the beginning and the end of the halt results in additional excitations on top of the KZ quasiparticles already excited near the critical point. However, the additional excitation energy is proportional to $\tau_Q^{-2}$, hence for large $\tau_Q$ it quickly becomes negligible when compared to the KZ excitation energy which is proportional to the density of excited quasiparticles and decays like $\tau_Q^{-1/2}$. Exact power spectra with and without the halt are compared in Fig. \ref{fig:halt} and they confirm this expectation. However, in experiment --- where $\tau_Q$ is limited --- instead of the sharp halt it may be more practical to avoid the discontinuities by performing a smooth ramp that simply takes longer to reach $g=0$ than the straight linear ramp. The ramp should just remain linear between $t_c\pm\hat t$ for the KZ scaling to remain unaffected. All of that being said, in the following we continue with the analytically convenient halt. 

An exact solution between initial $g=\infty$ and $g_w$ is the same as the one for the linear ramp, see (\ref{solution}). Then for time $t_w$ the Bogoliubov modes continue their evolution with a static Bogoliubov-de Gennes Hamiltonian at $g_w$. This stage is described by Eq. \eqref{dynBdG} with constant $g(t)=g_w$. Further evolution after the end of the halt is described by the general solution in (\ref{general}) but with time $t$ replaced by $t-t_w$. Its coefficients $a$ and $b$ are determined by matching this general solution with the $u_k,v_k$ at the end of the halt. This solution is continued until $g=0$, which is arrived at $t=t_w$, where the kink-kink correlator \eqref{CRpm} is measured. 

The dephasing time can be estimated based on the quasiparticle dispersion relation \eqref{epsilonk}. According to the excitation probability \eqref{LZpk} quasiparticles are excited up to small quasimomenta with 
\be 
\hat k^2=1/2\pi\tau_Q, 
\ee 
hence at $g=g_w$ the dispersion \eqref{epsilonk} can be approximated by
\bea  
\epsilon_k &\approx & 2\sqrt{(g_w-1)^2+g_w k^2} \nonumber\\
           &\approx & 2|1-g_w| +\frac{g_wk^2}{|1-g_w|}.
\eea
The last form is accurate when $2\pi\tau_Q\gg g_w (1-g_w)^{-2}$, i. e., either for slow enough quenches or deep enough in the ferromagnetic phase. A difference between dynamical phases, $2\epsilon_k t_w$, for $k=0$ and $k=\hat k$ becomes ${\cal O}(1)$ after a dephasing time 
\bea 
t_D \approx \frac{|1-g_w|}{g_w} \tau_Q.
\label{tD}
\eea 
Increasing the waiting time $t_w$ beyond $t_D$ should begin to have a noticeable effect on the anomalous correlator $\delta\beta_R$ eventually suppressing its magnitude to zero.

This rough estimate can be elevated to an accurate prediction. In Fig. \ref{fig:halt} we have shown that quasiparticle spectrum, $p_k$, does not depend on the waiting time. The dynamical phase $\varphi_k$ in \eqref{deltabeta} acquires an extra term, $\delta\varphi_k$, such that
\bea  
\delta\varphi_k -\delta\varphi_0 = 
2\epsilon_k t_w -2\epsilon_0 t_w = 
\frac{2g_w}{|1-g_w|} k^2 t_w .
\eea  
Just as the bare $\varphi_k$ in \eqref{quadraticvarphi} the extra term is quadratic in $k$. Consequently, new $\vert\delta\beta_R\vert$ for the ramp with a halt is obtained from the old $\vert\delta\beta_R\vert$ without a halt by a simple replacement:
$ 
\tau_Q \ln\tau_Q   \to  \tau_Q \ln\tau_Q + \frac{2g_w}{|1-g_w|} t_w.
$ 
Therefore, sole effect of the halt on the kink-kink correlator \eqref{scaledCR} is to replace the length scale $l$ in \eqref{l} with
\be 
l_w =
\hat\xi 
\sqrt{1+\left( \frac{3\ln\tau_Q+\frac{6g_w}{|1-g_w|} \frac{t_w}{\tau_Q}}{4\pi}  \right)^2}.
\label{lw}
\ee 
With the replacement the correlator after the halt becomes 
\be 
n^{-2} C^{KK}_R = 
\alpha~
\frac{\hat\xi}{l_w}
\left(\frac{R}{l_w}\right)^2 
e^{ -3\pi (R/l_w)^2 } 
- e^{ - 2\pi (R/\hat\xi)^2 }.
\label{scaledCRw}
\ee 
Here $l_w$ is longer than the bare $l$ corresponding to zero waiting time. Comparing \eqref{l} with \eqref{lw} we can infer that the waiting time begins to have noticeable effect when 
\be 
\frac{6g_w}{|1-g_w|} \frac{t_w}{\tau_Q} = 4\pi
\ee 
or, equivalently, for $t_w$ longer than a dephasing time
\be 
t_D=
\frac{2\pi}{3} 
\frac{|1-g_w|}{g_w}\tau_Q.
\ee 
This is when not only the correlation range, $l_w$, begins to increase but also the maximal value of the first term in \eqref{scaledCRw}, proportional to $\hat\xi/l_w$ and achieved at $R=l_w/\sqrt{\pi}$, begins to shrink. For $t_w\gg t_D$ this magnitude becomes negligible and the kink-kink correlator simplifies to the single anti-bunching term in \eqref{CRpmw}.

\section{ Higher order kink correlators }
\label{sec:higher}
The dephasing makes higher order correlators tractable. A connected $(M+1)$-point correlator reads
\be 
C_{R_1,\dots,R_M}=
\langle
K_0 K_{R_1} \dots K_{R_M}
\rangle_{c} .
\ee 
Thanks to permutation symmetry and translational invariance, we can assume $0=R_0<R_1<\dots<R_M$ without loss of generality. Expressing the kink number operators \eqref{K} with the Jordan-Wigner fermions \eqref{JW} allows us to write
\bea
&&
C_{R_1,\dots,R_M} = \nonumber\\
&&
(-2)^{-(M+1)} \times \nonumber\\
&&
\langle 
(b_0 a_1) ~ (b_{R_1} a_{R_1+1}) \dots (b_{R_M} a_{R_M+1})
\rangle_{c}.
\label{Cc}
\eea  
Here $b_n=c_n^\dag-c_n$ and $a_n=c_n^\dag+c_n$ are Majorana fermions. 
Their quadratic correlators are
\bea 
\langle b_m a_n \rangle &=&
\delta_{m,n}-2\alpha_{n-m}+2~{\rm Re}~\beta_{n-m},\\
\langle a_m a_n \rangle &=&
\langle b_m b_n \rangle  =
\delta_{m,n}+2i~{\rm Im}~\beta_{m-n}.
\eea 
After the dephasing $\beta_R\approx \frac14 {\rm sign}(R) \delta_{|R|,1}$ and, when $R_i$'s differ from each other by more than $2$, the relevant correlators simplify to
\bea 
\langle b_m a_n \rangle = -2\alpha_{n-m},~~
\langle a_m a_n \rangle = \langle b_m b_n \rangle  = 0.
\eea 
Given that $\alpha_R$ is even in $R$, the correlator becomes
\bea 
&& 
C_{R_1,R_2,\dots,R_M} \approx \nonumber \\
&&
\sum_{\{ i_1,\dots,i_M  \}}
\epsilon^{ i_0 i_1 \dots i_M }_{ i_1 \dots i_M i_0}~
\alpha_{R_{i_0}-R_{i_1}} \alpha_{R_{i_1}-R_{i_2}} \dots \alpha_{R_{i_M}-R_{i_0}} = \nonumber\\
&&
(-1)^M
\sum_{\{ i_1,\dots,i_M  \}}
\alpha_{R_{i_0}-R_{i_1}} \alpha_{R_{i_1}-R_{i_2}} \dots \alpha_{R_{i_M}-R_{i_0}}
.
\label{CRM}
\eea 
Here $i_0=0$ is fixed and the sum runs over all permutations of the set $\{ 1,\dots,M  \}$. For the assumed $\hat\xi\gg1$ we used $\alpha_{R\pm1}\approx\alpha_R$ as usual. 

For $M=1$ we recover \eqref{CRpmw} as expected. For $M=2$ we obtain a connected $3$-kink correlator:
\bea  
&&
n^{-3} C_{R_1,R_2} \approx \nonumber\\
&&
2 n^{-3} \alpha_{R_0-R_1} \alpha_{R_1-R_2} \alpha_{R_2-R_0} = \nonumber\\
&&
2 e^{-\pi[(R_0-R_1)/\hat\xi]^2} e^{-\pi[(R_1-R_2)/\hat\xi]^2} e^{-\pi[(R_2-R_0)/\hat\xi]^2}.
\eea 
While the negative kink-kink correlator \eqref{CRpmw} keeps pairs of kinks apart, the genuine three-body correlations turn out to be attractive. In general, odd/even kink correlations are attractive/repulsive.

\section{ Spin-spin correlator }
\label{sec:spinspin}

The dephasing also simplifies the ferromagnetic spin-spin correlator
\be 
C^{zz}_R =
\langle \sigma^z_n \sigma^z_{n+R} \rangle - 
\langle \sigma^z_n \rangle \langle \sigma^z_{n+R} \rangle.
\ee 
Given that $\langle \sigma^z_n \rangle=0$ for symmetry reasons, expressing the spin operators with the Jordan-Wigner fermions \eqref{JW} allows us to write
\be 
C^{zz}_R =
\langle 
b_0a_1b_1a_2\dotsb_{R-1}a_R
\rangle.
\ee 
In a similar way as for the higher order kink correlators, after dephasing the correlator becomes 
\be 
C^{zz}_R = {\rm Det} {\cal T}.
\ee 
Here 
$ 
{\cal T}_{ij}=\langle b_ia_{j+1} \rangle\approx \delta_{1+j-i,0}-2\alpha_{1+j-i}
$ 
for $i,j=1,\dots,R$. As ${\cal T}$ is a Toeplitz matrix, asymptotic behavior of the determinant for large $R$ can be obtained using standard methods \cite{schogo1} as
\be 
C^{zz}_R \propto 
\exp\left( -1.54 \frac{R}{\hat\xi} \right) 
\cos\left( \sqrt{ 4\pi \ln2 }~\frac{R}{\hat\xi} - \varphi_0 \right).
\ee 
This formula is the same as in Ref. \onlinecite{QKZteor-e} but here its range of applicability is much wider thanks to the extra dephasing during the waiting time $t_w$: $\ln\tau_Q\gg1$, that was assumed in Ref. \onlinecite{QKZteor-e}, is no longer required to make $l_w$ long enough, see \eqref{lw}. The correlator exhibits damped oscillations in function of $R/\hat\xi$. At short distance the ...-kink-kink-... train appears to have crystalline order. This is consistent with the anti-bunching seen in the kink-kink correlator: subsequent kinks keep safe distance from each other.

\section{Summary and discussion}
\label{sec:summary}

The connected kink-kink correlator at the end of the linear ramp is a sum of two terms. One of them is universal. It depends only on the spectrum of quasiparticles excited in the KZ regime, between $t_c\mp\hat t$ near the critical point, that depends only on the slope of the ramp at the critical point and the universal critical exponents. The other term is non-universal. It depends on how long the system is dragged across the ferromagnetic phase where the excited quasiparticles are dephasing by accumulation of quasimomentum-dependent dynamical phases. When the ramp is slowed on the way between the critical point and the zero transverse field for a waiting time much longer than $\tau_Q$ then the magnitude of the non-universal term begins to be suppressed by dephasing.

The dephased kink-kink correlator exhibits strong anti-bunching of kinks. Subsequent kinks along the chain are not allowed to approach each other closer than half of the typical distance between kinks. With dephasing it is possible to obtain also higher order kink correlators. The odd/even correlators turn out to be attractive/repulsive. The same dephasing makes the ferromagnetic spin-spin correlator tractable. In function of a distance it exhibits exponentially damped oscillations. The oscillations are consistent with the anti-bunching seen in the kink-kink correlator. 

The quantum Ising chain is integrable by mapping to non-interacting quasiparticles. One might wonder what happens when the Hamiltonian is supplemented with a perturbation breaking the integrability. Assuming thermalization, we can attempt a crude estimate by equating the final excitation energy per site in the Ising chain at $g=0$, which is $2n$ as each kink has energy $2$, with energy in thermal equilibrium at inverse temperature $\beta$, which is $1-\tanh\beta$ according to Ising's solution. For small density of kinks, $n$, we obtain $e^{-2\beta}\approx n$. Then the spin-spin correlation function after the equilibration becomes an exponent, $C^{zz}_R=\left(\tanh\beta\right)^R\approx(1-2n)^R$, with a correlation length $\xi\approx 1/2n=\hat\xi/2$. This is a unique scale of length in the equilibrium thermal state, the distinction between $\hat\xi$ and $l$ being washed out by the thermalization. 

Finally, a classical version of the Ising chain with Glauber dynamics should be mentioned as another similar toy model where KZM physics can be explored \cite{IsingGlauber_Krapivsky,IsingGlauber_Adolfo}. In this setting it may be also possible to obtain either kink correlations or closely related distribution of domain sizes in a similar way as for phase ordering kinetics \cite{IsingGlauber_Kinks}.

\acknowledgments
This research was supported in part by the National Science Centre (NCN), Poland together with the European Union through QuantERA ERA NET program No.~2017/25/Z/ST2/03028.

%
\bibliographystyle{apsrev4-1}
\bibliography{KZref.bib}
%
\end{document}